\documentclass[11pt]{article}
\usepackage{moriond}

\def\be{\begin{equation}}
\def\ee{\end{equation}}
\def\bea{\begin{eqnarray}}
\def\eea{\end{eqnarray}}

\def\lesssim{\ \hbox{\raise 2pt \hbox{$<$} \kern -13pt
                     \lower 3pt \hbox{$\sim$}}\ }
\def\greatersim{\ \hbox{\raise 2pt \hbox{$>$} \kern -13pt
                     \lower 3pt \hbox{$\sim$}}\ }

\def\pythia{{\sc Pythia}}
\def\herwig{{\sc Herwig}}
\def\powheg{{\sc Powheg}}

\begin{document}
\vspace*{4cm}
\title{Nonperturbative  corrections  and showering   in NLO-matched event  
generators~\footnote{Contributed at the XLVIII Rencontres de Moriond, March 2013.}}

\author{ S.~Dooling$^1$,   P.~Gunnellini$^1$,   F.~Hautmann$^{2,3}$  and  H.~Jung$^{1,4}$ }

\address{$^1$Deutsches Elektronen Synchrotron, D-22603 Hamburg\\
$^2$Theoretical Physics  Department, University of Oxford, 
Oxford OX1 3NP\\
$^3$Physics  and   Astronomy, University of  Sussex, 
Brighton BN1 9QH\\  
$^4$Elementaire Deeltjes Fysica, Universiteit Antwerpen, B 2020 Antwerpen
}

\maketitle\abstracts{We study  contributions   
 from  nonperturbative effects and parton showering  in NLO event generators,   
and present  applications to jet final states.  We find $p_T$-dependent  and 
rapidity-dependent   corrections  which can affect  
 the shape of observed jet distributions at the  LHC. 
We illustrate numerically the  kinematic shifts 
in longitudinal momentum distributions   from 
the implementation of energy-momentum conservation in collinear shower algorithms.}   

Monte Carlo event generators are used in  analyses  of complex final states 
at the Large Hadron Collider (LHC)~\cite{hoeche11} 
 both to supplement finite-order perturbative calculations with  all-order 
  QCD radiative terms,   encoded by parton showers,   
  and to incorporate 
nonperturbative effects from hadronization, multiple  parton interactions, 
underlying events~\cite{pythref,herwref}. In this article we report results 
from   our 
study~\cite{sampao}   of  
 nonperturbative (NP) and parton-showering (PS) 
corrections in the context of matched NLO-shower Monte Carlo generators. 
The results we  present refer to  jet final states. Further results for    
massive  states  may be found in~\cite{sampao}.

LHC experiments have measured 
 inclusive jet production~\cite{atlas-1112,CMS:2011ab}  
 over  a     kinematic  range in transverse momentum 
and rapidity  much larger  than in any previous collider experiment. 
Baseline comparisons  
 with Standard Model theoretical predictions   are   based 
 either  on 
next-to-leading-order (NLO) QCD calculations, supplemented with 
nonperturbative  (NP) corrections estimated from 
Monte Carlo  event generators~\cite{atlas-1112,CMS:2011ab},  or on NLO-matched parton shower event 
generators~\cite{ma}. 
The first kind of comparison shows   
 that the NLO 
calculation agrees with data at central rapidities, while increasing deviations are 
seen with increasing rapidity at large transverse momentum $p_T$~\cite{atlas-1112}.   
The question  arises  of whether such    behavior is associated with 
 higher-order   perturbative contributions or with nonperturbative components 
 of the cross section.  The second kind of comparison,  
 based on  \powheg\  calculations~\cite{alioli}      
in which  NLO matrix elements are  matched with parton showers~\cite{pythref,herwref}, 
 improves the description of data, 
indicating that 
higher-order radiative  contributions  taken into account via 
parton showers are  
numerically  important.  At the same time, the results  show large differences between 
\powheg\  calculations interfaced with different shower generators, \pythia~\cite{pythref}  
 and \herwig~\cite{herwref}, in the forward rapidity region, pointing to enhanced       
 sensitivity to  details of the showering. 

NP  correction factors are obtained 
 in~\cite{atlas-1112,CMS:2011ab}    by using 
 leading-order  Monte Carlo (LO-MC) generators~\cite{pythref,herwref}. 
The method to determine these factors  is to compare   
  a  Monte Carlo simulation including   
parton showers, multiparton interactions and hadronization,  
and   a  Monte Carlo simulation including  only 
 parton showers     in addition to the LO hard process. 
While this  is a natural way to 
estimate NP   corrections from LO+PS event generators,  it is noted in~\cite{sampao} that
 when these  corrections  are combined   
 with NLO parton-level results a potential 
   inconsistency arises  because the 
    radiative correction  from the first  gluon emission   is treated  at different levels  of 
    accuracy     in the two parts of the  calculation.    To avoid this, 
Ref.~\cite{sampao}   proposes a method 
which  uses 
NLO Monte Carlo (NLO-MC) generators to determine the correction. 
In this case one can consistently  assign correction factors 
to be applied to  NLO calculations.  This  method   allows one to 
study  separately 
 correction factors to the  fixed-order calculation  due to  parton showering  
effects. 
To  do this,  Ref.~\cite{sampao}    introduces  
  the   nonperturbative (NP) and showering (PS)  correction factors,  $K^{NP}$  and  $K^{PS}$, 
 as 
\begin{equation} 
\label{npK2} 
K^{NP} =  { {  N_{NLO-MC}^{(ps+mpi+had)} } /  {  N_{NLO-MC}^{(ps)} }}   \;\;   , 
\end{equation} 
\begin{equation} 
\label{npK3} 
K^{PS} =  { {  N_{NLO-MC}^{(ps)} } /  {  N_{NLO-MC}^{(0)} }}   \;\;   ,  
\end{equation}
where 
$(ps+mpi+had)$ denotes  a  simulation including   
parton showers, multiparton interactions and hadronization,  while 
$(ps)$ denotes a simulation including   
parton showers only.      
    The  denominator  in Eq.~(\ref{npK3})   is  defined  by switching off 
all components beyond NLO in the Monte Carlo simulation. 
\vspace*{-0.5cm} 
\begin{figure}[htbp]
\begin{center}
\includegraphics[scale=.24]{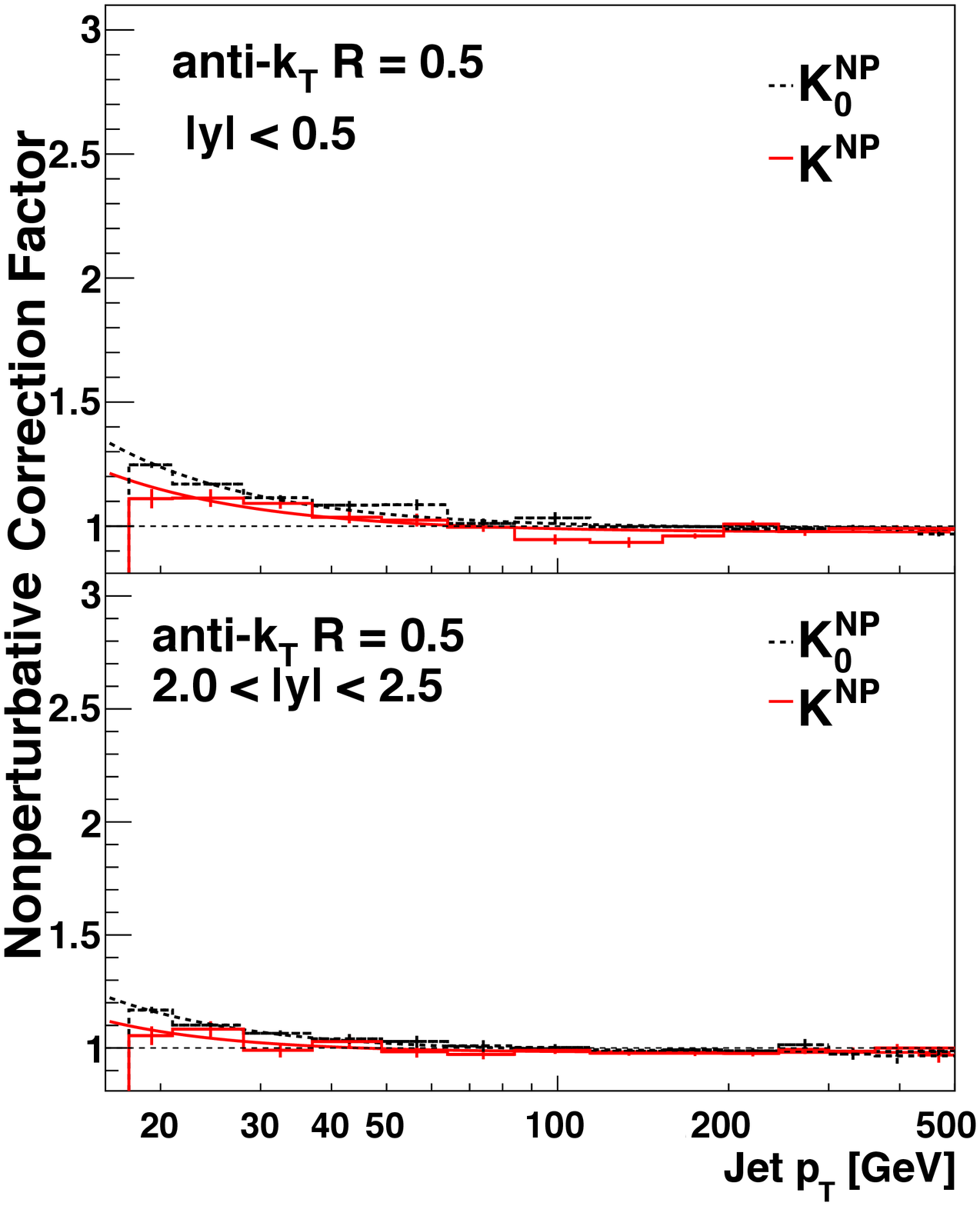}
\includegraphics[scale=.24]{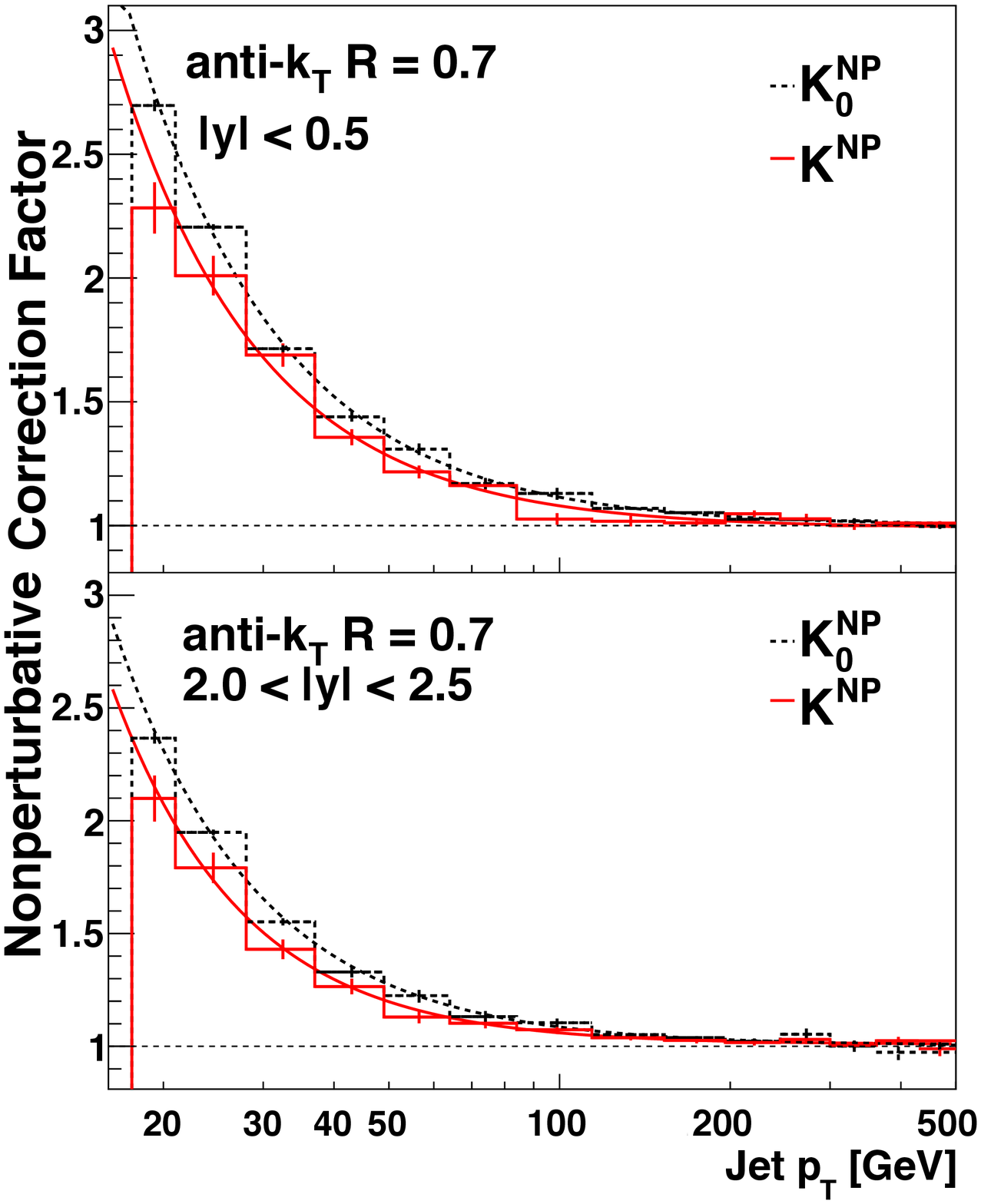}
\caption{\it The NP  correction factors to jet transverse 
momentum distributions 
 obtained 
 using  \protect\pythia\  and \protect\powheg\  respectively, 
 for $|y|<0.5$ and $2 < |y|< 2.5$.  
Left: $R=0.5$; Right:  $R=0.7$.  
}  
\label{fig:np1}
\end{center}
\end{figure}

\begin{figure}[htbp]
\begin{center}
\includegraphics[scale=.24]{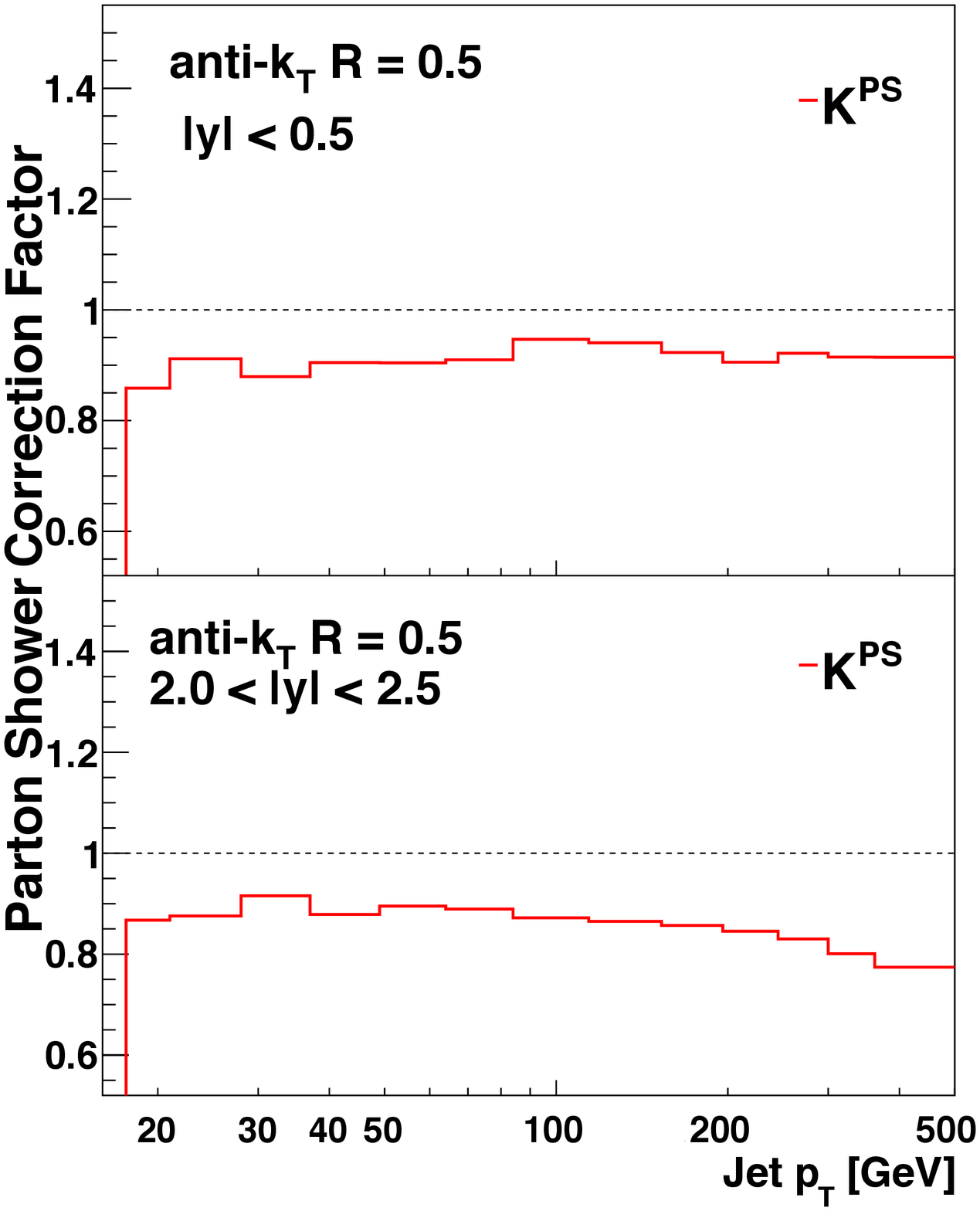}
\includegraphics[scale=.24]{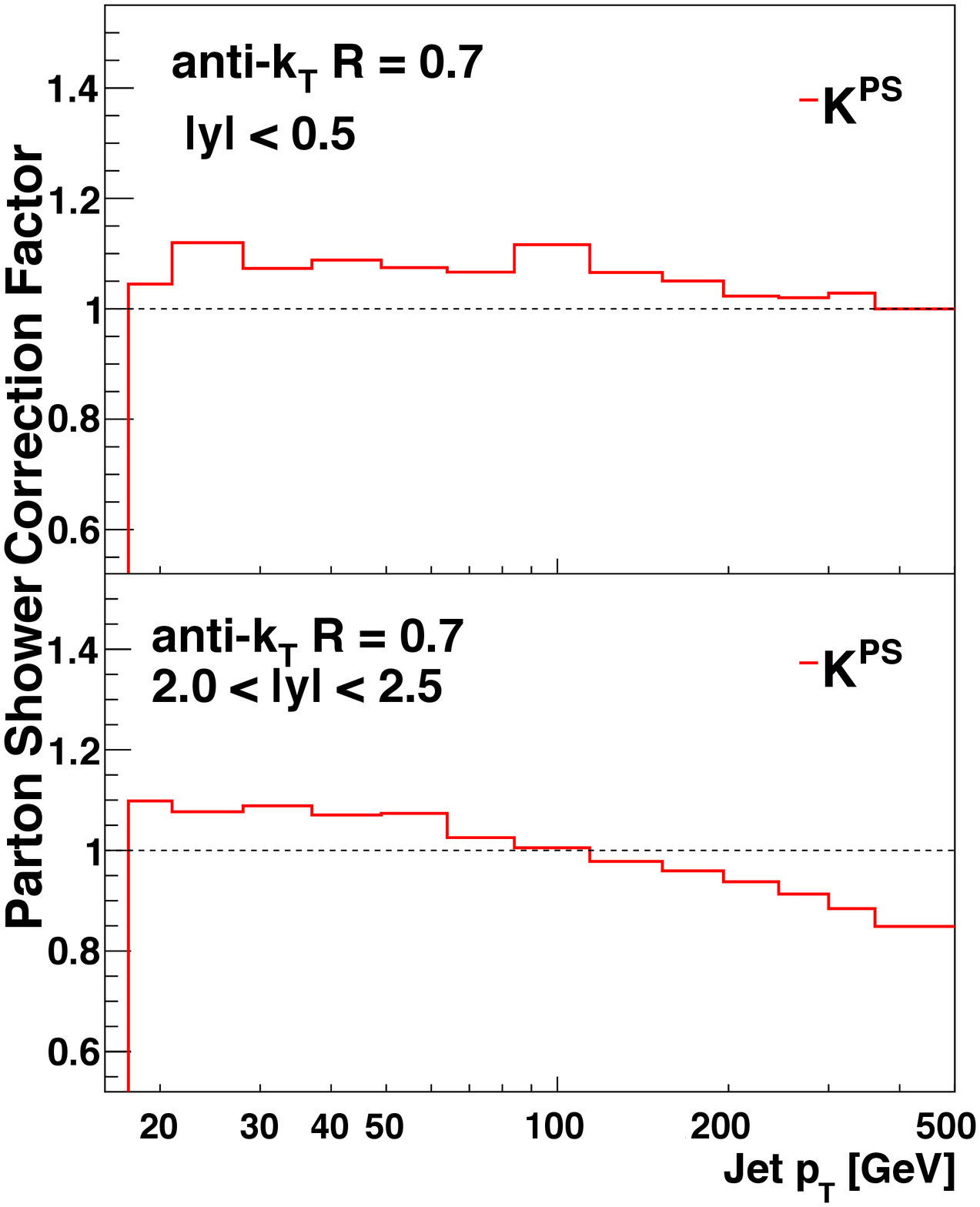}
\caption{\it The  parton shower
 correction factor to jet transverse momentum  distributions, obtained from  
 Eq.~(\ref{npK3})  using 
 \protect\powheg\   
   for $|y|<0.5$ and $2 < |y|< 2.5$.
Left: $R=0.5$; Right:  $R=0.7$.  
}  
\label{fig:np2}
\end{center}
\end{figure}

The factor $ K^{NP} $ in  Eq.~(\ref{npK2}) 
differs from the LO-MC NP factor~\cite{atlas-1112,CMS:2011ab} 
  because of the different definition 
of the hard process. In particular the multi-parton interaction p$_T$ cut-off scale is 
different in the LO and NLO cases. 
Numerical results are shown in  Fig.~\ref{fig:np1}. 
The factor $ K^{PS} $ in  Eq.~(\ref{npK3}), on the other hand,   
  is new. It singles out 
contributions due to parton showering and has not been considered in previous analyses. 
Unlike the NP correction,  it gives  in general finite effects also at large p$_T$. 
Results are  plotted  in 
 Fig.~\ref{fig:np2},  showing  that 
 this  correction 
  is $y$  and p$_T$ dependent, especially when rapidity is 
non-central, so that it cannot be treated as a rescaling.

The correction factor in  Fig.~\ref{fig:np2} comes from initial-state and final-state showers.  
These are   interrelated so  that the 
      combined effect is nontrivial and is not   obtained 
     by simply adding the two~\cite{sampao}.   
 The effect from parton shower is largest at large $|y|$, where  the 
initial-state parton shower is mainly contributing 
at low $p_T$, while the final-state parton shower is contributing significantly 
 over the whole $p_T$ range.   

  The main  effect of  initial-state  showering   is associated with the 
 kinematic  shifts  in   longitudinal momentum  
 distributions first noted in~\cite{coki-1209}. 
 These shifts  result, quite generally, from   combining   the approximation 
of collinear, on-shell partons with the  requirements  of  energy-momentum conservation in the 
Monte Carlo generator. More precisely, 
the Monte Carlo  first generates hard subprocess 
 events  in which   the momenta   $k_j$   of the 
partons initiating the hard scatter are on shell, and are taken to be fully collinear 
with the incoming  state  momenta. 
 Next  the 
showering algorithm is applied, and complete final states are generated including 
additional QCD radiation from the initial  and final  parton cascades. 
As a result of QCD  showering, the momenta $k_j$   are no longer exactly  collinear.   
Their transverse momentum is   to be   compensated  by a change  in the 
kinematics of the hard scattering subprocess. 
By  energy-momentum conservation, however,   this  implies  a 
reshuffling, event by event,  in  the  
fractions $x_j$  of longitudinal momentum 
 carried by  the partons  scattering off each other in the hard  subprocess. 

\begin{figure}[htbp]
\begin{center}
\includegraphics[scale=.5]{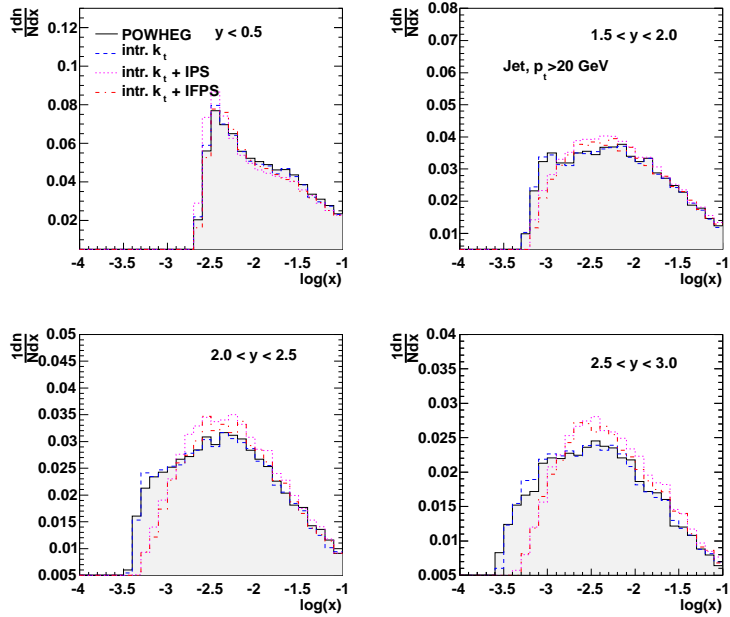}
\caption{\it Distributions in the parton longitudinal momentum fraction $x$ 
 before (POWHEG) and after parton showering (POWHEG+PS),  
for  inclusive jet  production  at  different rapidities for jets with $p_T> 18 $ GeV obtained 
by  the anti-kt jet  algorithm   
with $R=0.5$.
Shown is the effect of intrinsic $k_t$, initial (IPS) and initial+final state (IFPS) parton shower.    
}  
\label{fig:fig1}
\end{center}
\end{figure}

The size of the shift is    illustrated in  Fig.~\ref{fig:fig1}~\cite{sampao} for the case of 
 jets  produced at different rapidities,  by comparing 
 the distribution in the parton longitudinal momentum fraction $x$ 
before  parton showering 
and after parton  showering.  We see that the longitudinal shift is negligible 
 for  central rapidities but becomes 
significant for $ y >1.5$.   It characterizes the highly asymmetric 
parton kinematics~\cite{jhep09}  which   becomes important  for the first time 
at the LHC  in significant regions of phase space.   
 Although Fig.~\ref{fig:fig1} is obtained using 
a  particular  NLO-shower  matching 
 scheme (\powheg),  the effect    is common  to any calculation matching NLO 
 with collinear showers.  
  On the other hand,   this is avoided  in shower algorithms 
   using transverse momentum dependent  
   parton  distributions~\cite{avsar11,unint09,muld-rog-rev,martin}  
     from the beginning,  
     as  for instance in~\cite{Jung:2010si,jadach09}.

 In summary, 
the nonperturbative correction factor $K^{NP}$ introduced from NLO-MC in Eq.~(\ref{npK2})   
gives non-negligible differences 
compared to 
the LO-MC contribution~\cite{atlas-1112,CMS:2011ab} 
at low to intermediate jet $p_T$, while 
  the showering  correction factor  $K^{PS}$  of  Eq.~(\ref{npK3})   
  gives  significant effects  over the whole $p_T$  range  and is largest at 
  large  jet  rapidities $y$.   
 Because of this     $y$ and $p_T$ dependence,   
    taking properly into account    NP and showering correction factors     
  changes  the  shape  of jet distributions, and   may  thus  influence   the comparison 
  of theory predictions with experimental data. 
Besides jets, longitudinal momentum  shifts as  in  Fig.~\ref{fig:fig1} 
  also affect massive 
final states~\cite{sampao}  such as Drell-Yan $ Z / W $ production. 
We anticipate   
that  the   showering correction factors     
  will    be     relevant   in particular   in    fits  for 
parton distribution functions   using  inclusive jet and vector boson data.

\section*{Acknowledgments}
We  are grateful to    the Moriond organizers and staff  for the invitation to 
 an exciting  conference, and  to  the Moriond   participants  for   interesting discussions.


\section*{References}

\begin{thebibliography}{99}


\bibitem{hoeche11} 
        See    S.~H{\" o}che,  SLAC  preprint SLAC-PUB-14498 (2011) for a recent review.  

\bibitem{pythref}
         T.~Sj{\" o}strand, S.~Mrenna and P.~Skands, JHEP 
         {\bf 0605} (2006) 026. 


\bibitem{herwref}
  G.~Corcella {\it et al.},
  JHEP {\bf 0101} (2001) 010
  [arXiv:hep-ph/0011363]; 
  G.~Corcella {\it et al.},
  arXiv:hep-ph/0210213.


\bibitem{sampao}
     S.~Dooling,  P.~Gunnellini, F.~Hautmann and 
     H.~Jung,    arXiv:1212.6164  [hep-ph]. 
  

\bibitem{atlas-1112} 
            ATLAS Coll. (G.~Aad et al.),      Phys.\  Rev.\ D {\bf  86}  (2012) 014022.  



\bibitem{CMS:2011ab}
           CMS Coll.  (S.~Chatrchyan et al.),    
                                          Phys.\ Rev.\ Lett.\  {\bf 107} (2011) 132001; 
            arXiv:1212.6660  [hep-ex]. 


\bibitem{ma} 
          P.~Nason and B.R.~Webber,    arXiv:1202.1251  [hep-ph]. 



\bibitem{alioli} 
        S.~Alioli et al.,       
         JHEP {\bf 1104} (2011)  081.   



   

\bibitem{coki-1209} 
              F.~Hautmann and H.~Jung,     
              Eur.\  Phys.\   J.\  C {\bf 72} (2012) 2254.       


\bibitem{antiktalgo}     
      M.~Cacciari, G.~Salam and G.~Soyez,     JHEP {\bf 0804} (2008) 063. 
 

\bibitem{jhep09} 
       M.~Deak et al.,           JHEP {\bf 0909} (2009) 121;   arXiv:0908.1870;     
       arXiv:1012.6037 [hep-ph];  
          Eur.\  Phys.\   J.\  C {\bf 72} (2012) 1982;  arXiv:1206.7090 [hep-ph]. 

\bibitem{avsar11}
     E.~Avsar,   arXiv:1203.1916  [hep-ph];      arXiv:1108.1181 [hep-ph]. 



\bibitem{unint09} 
    F.~Hautmann,   
     Acta  Phys.\  Polon.\  B  {\bf 40} (2009) 2139;  Phys.\ Lett.\  B {\bf  655} (2007) 26; 
            F.~Hautmann and H.~Jung,  
     arXiv:0712.0568  [hep-ph];        arXiv:0805.1049  [hep-ph];    
        arXiv:0808.0873. 
     

\bibitem{muld-rog-rev} 
         P.J.~Mulders,         Pramana {\bf 72} (2009)  83; 
             P.J.~Mulders and T.C.~Rogers,   arXiv:1102.4569  [hep-ph].  

\bibitem{martin} 
        F.~Hautmann,  M.~Hentschinski and H.~Jung, 
          Nucl. Phys.  B{\bf  865} (2012) 54;     
       arXiv:1205.6358  [hep-ph];    arXiv:1209.6305  [hep-ph].  


 \bibitem{Jung:2010si}     
       H.~Jung et al.,      Eur.\  Phys.\   J.\  C {\bf 70} (2010) 1237;    arXiv:1206.1796  [hep-ph].   
            

  \bibitem{jadach09}     
        S.~Jadach   and  M.~Skrzypek, 
            Acta  Phys.\  Polon.\  B  {\bf 40} (2009) 2071; 
             S.~Jadach, M.~Jezabek,  A.~Kusina, W.~Placzek    and  M.~Skrzypek,   
    arXiv:1209.4291 [hep-ph].     


\end{thebibliography}
\end{document}